\documentclass[aps,prl,twocolumn,superscriptaddress,amsmath,amssymb]{revtex4-2}

\usepackage{graphicx}
\usepackage{dcolumn}
\usepackage{bm}
\usepackage{hyperref}
\usepackage{xcolor}

\begin{document}

\title{Chemical Medium-Range Order Enables Stoichiometric Rigidity}

\author{Kejun Liu}
\email{kjliu@suda.edu.cn}
\affiliation{State Key Laboratory of Bioinspired Interface Material Science, Institute of Nano \& Functional Materials, Soochow University, Suzhou 215123, China}

\date{\today}

\begin{abstract}
Maxwell counting predicts an isostatic threshold at $\langle r\rangle = 2.4$ for covalent network glasses, but which structural correlations actually produce rigidity near this point is still unclear. In this work, we test four candidates: enthalpic stress, chemical defects, geometric interlocking, and medium-range order (MRO). We use a locally tree-like configuration model as a zero-MRO baseline and apply perturbations to test each candidate. We find that (i)~enthalpic stress \emph{delays} rigidity rather than enabling it; (ii)~chemical defects require fractions (${\sim}40\%$) far above experimental values (${\sim}16\%$ in GeSe$_2$); (iii)~geometric linking density does not govern the threshold location, which is instead set by loop-induced redundancy; and (iv)~only phenomenological MRO proxies recover rigidity at experimentally accessible strengths. Consequently, chalcogenide intermediate-phase data and amorphous SiO$_2$ ring statistics positively implicate chemical MRO, while DNA spatial networks independently rule out pure geometric entanglement. We conclude that rigidity near the Maxwell threshold requires chemistry-specific correlations beyond pure connectivity.
\end{abstract} 

\maketitle

\begin{figure*}[t]
  \centering
  \includegraphics[width=\textwidth]{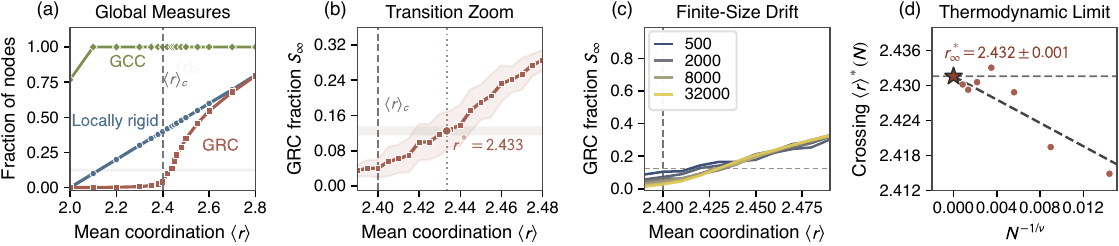}
  \caption{\label{fig:baseline} Configuration-model baseline establishing the zero-MRO reference frame. (a)~Global rigidity measures vs.\ mean coordination: locally rigid fraction, giant rigid component (GRC), and connectivity giant component, with the Maxwell point at $\langle r\rangle_c = 2.4$. (b)~Transition zoom showing the GRC crossing the $S_\infty = 12.5 \pm 1\%$ reference window at $\langle r\rangle^* \approx 2.433$ ($N=5000$, 20 trials). (c)~Finite-size drift of $S_\infty$ for $N \in [500, 32\,000]$. (d)~Thermodynamic-limit extrapolation of $\langle r\rangle^*(N)$ vs.\ $N^{-1/\nu}$ yields $\langle r\rangle^*_\infty = 2.431 \pm 0.002$, agreeing with the mean-field prediction $2.428$ within $0.15\%$.}
\end{figure*}

The structural origin of mechanical rigidity in glasses has been debated since Zachariasen's random-network model~\cite{zachariasen1932}. Phillips~\cite{phillips1979} and Thorpe~\cite{thorpe1983} later proposed that rigidity emerges at the Maxwell isostatic point $\langle r\rangle_c = 2.4$. In mean-field theory, a network with perfect chemical order---rigid ($k\geq 3$) and floppy ($k=2$) units fully segregated---should remain globally floppy because rigid clusters are isolated by floppy chains. This is a connectivity effect, distinct from thermodynamic phase separation. The rigid and floppy subnetworks therefore remain connectivity-isolated, an observation implicit in the rigid-backbone picture of Jacobs and Thorpe~\cite{jacobs1995,thorpe1983} that we here make analytically exact on a zero-MRO baseline. The puzzle is then: why are real stoichiometric glasses mechanically stable near $\langle r\rangle_c = 2.4$ despite this connectivity obstruction? Experiments clearly show rigidity, but it has been difficult to identify which correlation is responsible---chemical ordering, medium-range motifs~\cite{micoulaut2022}, or finite-dimensional effects~\cite{chubynsky2007}---because glass structure lacks atomic-scale resolution~\cite{micoulaut2023review}.

Rigidity percolation has recently expanded to polymer gels~\cite{broedersz2014}, colloidal assemblies~\cite{zaccone2019correlated,luo2019direct}, DNA networks~\cite{nishikawa2022rigidity}, and limited-valence gels via m-percolation theory~\cite{dias2025}. DNA dendrimers~\cite{neophytou2024} are especially useful because Watson-Crick pairing gives covalent-analog connectivity, while ring interlocking tests purely geometric effects. The same question unites all these systems: what does connectivity topology alone produce, and what needs chemistry-specific correlations?

Although ring-based MRO has been invoked heuristically in earlier theoretical work~\cite{micoulaut2003,micoulaut2022,micoulaut2023review}, a quantitative baseline that separates connectivity-level isolation from MRO and allows each candidate mechanism to be independently falsified has been lacking. To provide such a baseline, we use the locally tree-like configuration model~\cite{molloy1995} as a zero-MRO reference. For $\{2,3\}$ networks (As$_2$Se$_3$-type), stoichiometry gives $\langle r\rangle = 2.4$; for $\{2,4\}$ networks (GeSe$_2$-type), we tune composition to $\langle r\rangle = 2.4$ (20\% rigid $k=4$, 80\% floppy $k=2$). Then we test four mechanisms that could bridge this connectivity obstruction: enthalpic stress, chemical defects, geometric linking, and phenomenological MRO. In this baseline, rigidity onset coincides exactly with the Maxwell point. Amorphous SiO$_2$ (chemical MRO without geometric linking) and DNA networks (geometric linking without chemical defects) provide independent benchmarks.

\noindent\textbf{Network model and local rigidity---} In our model, the network is a configuration-model graph~\cite{molloy1995} with $N$ nodes and a two-point degree distribution. In 3D Phillips-Thorpe theory, a node with coordination $k_i$ has $C_i = k_i/2 + \max(0,\, 2k_i - 3)$ constraints. It is locally rigid when $C_i \geq 3$, that is, $k_i \geq 3$. On tree-like graphs, this criterion matches the pebble-game result: no redundant edges are detected (SM, Sec.~S7), which is expected because short loops are absent. We define the giant rigid component (GRC) as the largest connected cluster of locally rigid nodes, with fraction $S_\infty$.

\noindent\textbf{Analytical framework---} The Maxwell count is $f_M(\langle r\rangle) = 6 - 5\langle r\rangle/2$~\cite{thorpe1983}, so the isostatic point is $\langle r\rangle_c = 2.4$. We set $\alpha = \langle r\rangle - 2$, so the degree and excess-degree generating functions become $g_0(x) = (1-\alpha)x^2 + \alpha x^3$ and $g_1(x) = [2(1-\alpha)x + 3\alpha x^2]/(2+\alpha)$. The branching probability $u$ that a bond does not reach the GRC obeys $u = 2(1-\alpha)/(2+\alpha) + [3\alpha/(2+\alpha)]u^2$ (SM, Sec.~S1), and its physical root is $u^* = 2(1-\alpha)/(3\alpha)$. A macroscopic GRC exists if and only if $u^* < 1$, which means $\alpha > 2/5$ or $\langle r\rangle > 2.4$. Consequently, in the mean-field (tree-like) limit, the geometric rigidity onset \emph{coincides exactly} with the Maxwell point. The backbone fraction is
\[
S_\infty = \alpha\!\left[1 - (u^*)^3\right],
\]
Setting $S_\infty = 12.5\%$ yields $\alpha^* \approx 0.4283$ and $\langle r\rangle^*_{\rm MF} \approx 2.428$ (SM, Sec.~S2). This 12.5\% level is also the classical ER site-percolation threshold $p_c = 1/8$ (SM, Sec.~S4), and it is robust because a $\pm 1\%$ window shifts $\langle r\rangle^*$ by only $\pm 0.0025$ (SM, Sec.~S5). In real networks, short loops create redundant constraints and can shift the rigidity threshold in either direction: upward when the redundancy is purely geometric, as in DNA spatial networks where $\langle k\rangle_c \approx 2.52$~\cite{jacobs1995,neophytou2024}, or downward when chemically directed medium-range order bridges rigid clusters, as in chalcogenide glasses~\cite{sartbaeva2007,zeidler2017}. The mean-field value therefore provides a clean reference for measuring that shift~\cite{chubynsky2007}. In addition, this degeneracy is universal: both $\{2,3\}$ and $\{2,4\}$ networks give $\langle r\rangle_c = 2.4$ (SM, Sec.~S1).

\noindent\textbf{Numerical results and finite-size scaling---} Figure~\ref{fig:baseline}(a) shows $S_\infty$ and the locally rigid fraction $p_r$ vs.\ $\langle r\rangle$ ($N=5000$, 20 trials). The GRC turns on at $\langle r\rangle_c = 2.4$ and grows continuously above it. Panel~(b) shows the transition zoom: using the 12.5\% reference level, simulations give $\langle r\rangle^* \approx 2.433$, consistent with the mean-field value $2.428$. Finite-size scaling for $N \in [500, 32\,000]$ (panels c--d) yields $\langle r\rangle^*_\infty = 2.431 \pm 0.002$ and a drift exponent $\nu = 1.47 \pm 0.43$ (SM, Sec.~S5). Here $\nu$ simply tracks how the crossing coordinate moves with system size. The extrapolated limit is insensitive to this uncertainty: varying $\nu$ across $(1.0, 1.9)$ shifts $\langle r\rangle^*_\infty$ by less than $0.001$ (SM, Sec.~S5). The extrapolated value agrees with the mean-field prediction to within $0.15\%$. However, with perfect chemical order at $\langle r\rangle = 2.4$, the model predicts a globally floppy state ($S_\infty = 0$): rigid and floppy nodes remain connectivity-isolated. This result contradicts the experimental fact that real glasses are rigid.

\begin{figure*}[t]
  \centering
  \includegraphics[width=\textwidth]{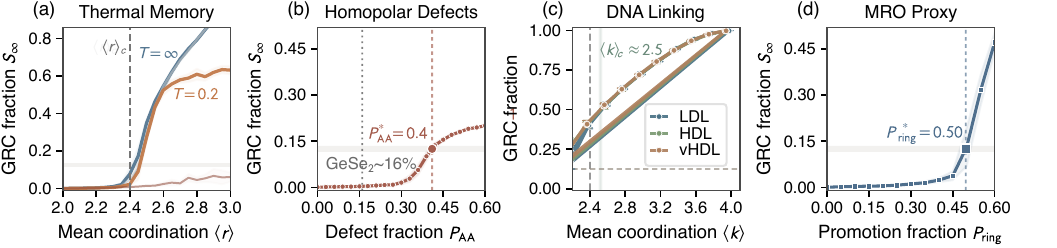}
  \caption{\label{fig:perturbations} Systematic hierarchy of rigidity mechanisms. (a)~Enthalpic stress test: Metropolis growth with rigid-rigid bond penalty ($\varepsilon=1$) at five temperatures [$T \in \{\infty, 1.0, 0.5, 0.2, 0.1\}$] shifts $\langle r\rangle^*$ rightward, from $2.425$ at $T=\infty$ to $2.459$ at $T=0.2$, widening the phase window rather than narrowing it; only $T=\infty$ and $T=0.2$ are labeled in the panel for readability. (b)~Chemical defects: homopolar $P_{\rm AA}$ bonds bridge the rigid clusters only at $P_{\rm AA}^* \approx 0.41$, far exceeding the experimental Se-Se-Se fraction (${\sim}16\%$ in GeSe$_2$). (c)~DNA linking: bond dilution across three phases yields a shared rigidity threshold $\langle k\rangle_c \approx 2.52 \pm 0.01$ despite 7-fold variation in linking density. (d)~Phenomenological MRO proxy. Amorphous SiO$_2$ structures show an average of 17 chemical rings per configuration, confirming that MRO in $\{2,4\}$ networks is ring-mediated~\cite{sharma1981,micoulaut2019}. In the tree-like configuration model, where such explicit rings are absent, we introduce $P_{\rm ring}$ as an effective proxy for their connectivity-level unblocking contribution; promoting fraction $P_{\rm ring}$ of B nodes to the rigid set restores rigidity at $P_{\rm ring}^* \approx 0.50$.}
\end{figure*}

\noindent\textbf{Candidate mechanisms for rigidity---} Which physical mechanism restores rigidity in chemically ordered glasses near the Maxwell threshold? In Fig.~\ref{fig:perturbations}, we test each candidate at experimentally accessible parameter values. Real Ge-Se glasses show narrower intermediate phases shifted below the mean-field baseline. A perturbation that moves the threshold downward is therefore physically relevant.

\textit{Test 1: Enthalpic stress [Fig.~\ref{fig:perturbations}(a)].---} We use Metropolis growth with $H = \varepsilon \sum_{\langle ij\rangle} \delta_{k_i \geq 3}\,\delta_{k_j \geq 3}$ to penalize rigid-rigid bonds ($T \in \{\infty, 1.0, 0.5, 0.2, 0.1\}$, $N=3000$, 8 trials). We find that lower temperature shifts $\langle r\rangle^*$ \emph{rightward} (from 2.425 at $T=\infty$ to 2.459 at $T=0.2$). Enthalpic stress thus widens the phase window rather than narrowing it. 

\textit{Test 2: Chemical defects [Fig.~\ref{fig:perturbations}(b)].---} At fixed composition $\langle r\rangle = 2.42$ (21\% $k=4$, 79\% $k=2$), we increase the homopolar fraction $P_{\rm AA}$ and find that it bridges isolated rigid clusters. The GRC rises from 0 to ${\sim}20\%$ with threshold $P_{\rm AA}^* = 0.41$. This value is well above the experimental Se-Se-Se fraction (${\sim}16\%$ in GeSe$_2$). 

\textit{Test 3: Pure topological linking [Fig.~\ref{fig:perturbations}(c)].---} We also study DNA spatial networks with mechanical ring interlocking~\cite{neophytou2024}. These networks vary 7-fold in linking density ($\rho_{\rm link} = 0.05$--$0.35$), but they share the same rigidity threshold ($\langle k\rangle_c = 2.52 \pm 0.01$; ANOVA shows no significant difference between groups, $p=0.95$). The shift from 2.4 to 2.52 thus reflects loop-induced redundancy, not linking density. 

\textit{Test 4: Phenomenological MRO [Fig.~\ref{fig:perturbations}(d)].---} Amorphous SiO$_2$ shows ${\sim}17$ chemical rings per structure, indicating that MRO in $\{2,4\}$ networks is ring-mediated~\cite{sharma1981,micoulaut2019,micoulaut2003}. In the tree-like model, explicit rings are absent, so we introduce $P_{\rm ring}$ as an effective proxy: promoting fraction $P_{\rm ring}$ of B nodes to the rigid set restores rigidity at $P_{\rm ring}^* \approx 0.50$. This upgrade is the mean-field image of ring closure: a B node belonging to a ring has one bond beyond its tree-like neighbors, giving it effective coordination $k\geq 3$ and the same constraint count as a primary rigid site (SM, Sec.~S9). Because bridging O atoms in $a$-SiO$_2$ participate in $\geq 6$-membered Si--O rings almost universally~\cite{sharma1981,micoulaut2003}, the experimental ring-participation fraction sits well above $P_{\rm ring}^* \approx 0.50$, whereas the defect threshold $P_{\rm AA}^* \approx 0.41$ lies far above the NMR-measured Se-Se-Se fraction of ${\sim}16\%$ in GeSe$_2$~\cite{gjersing2010}. Medium-range reorganization is thus the only mechanism that crosses its threshold within experimentally realized parameter ranges~\cite{micoulaut2022,tavanti2020,tsiulyanu2022,sidebottom2025}.

\noindent\textbf{Computational comparisons---} To benchmark the hierarchy, we compare two independent simulations: amorphous SiO$_2$ (chemical MRO, no geometric linking) and DNA dendrimers (geometric linking, no chemical defects). Both were analyzed with the same ring/linking engine, which is adapted from Neophytou et al.~\cite{neophytou2024}; details are in the SM.

\textit{SiO$_2$ oxide glasses.---} DFT-AIMD simulations of 30 SiO$_2$ structures (Materials Project MPMorph workflow~\cite{materialsproject}) contain chemical rings but \emph{no geometric linking pairs}. This result confirms that oxide MRO is purely chemical.

\textit{DNA spatial networks.---} In DNA dendrimers~\cite{neophytou2024}, geometric linking raises the Fiedler eigenvalue $\lambda_2$ by 72\% but leaves the rigidity threshold unchanged. Because DNA differs from inorganic glasses in bonding, solvent, and flexibility, this comparison tests topology rather than chemistry.

These two benchmarks agree: chemical MRO shifts the threshold, while geometric linking only stiffens the spectrum. $P_{\rm ring}$ therefore parametrizes chemically mediated constraints (ring-closure motifs in Ge-Se or Si-O), not topological entanglement.

\noindent\textbf{Experimental context---} Boolchand and coworkers~\cite{boolchand2001,boolchand2005} identified narrow stress-free intermediate phases in Ge-Se and As-Se glasses near stoichiometry. This feature cannot be explained by mean-field theory alone~\cite{thorpe2000,zeidler2017}. Our marker $\langle r\rangle^* = 2.431$ falls inside the reported IP window $[2.35, 2.47]$~\cite{zeidler2017}, and the vanishing non-reversing enthalpy ($\Delta H_{nr} \to 0$) marks the intermediate phase~\cite{boolchand2001}. In addition, $^{77}$Se NMR finds only ${\sim}16\%$ homopolar Se-Se-Se in GeSe$_2$~\cite{gjersing2010}, which is far below the defect threshold $P_{\rm AA}^* \approx 0.41$. For SiO$_2$, Raman studies reveal ring-based MRO motifs~\cite{sharma1981}. For DNA networks, valence-controlled mechanics and topological linking have been observed~\cite{conrad2019,palombo2025}. All three experimental systems converge on the same conclusion: chemical MRO, not defects or topology alone, stabilizes rigidity.

\noindent\textbf{Discussion---} The tree-like baseline fixes the GRC onset at $\langle r\rangle_c = 2.4$ and places the finite-size-extrapolated marker at $\langle r\rangle^* = 2.431 \pm 0.002$. These tests establish a mechanistic hierarchy: chemical MRO ($P_{\rm ring}^* \approx 50\%$) is necessary and sufficient to connect rigid clusters through the floppy matrix, whereas geometric linking enhances stiffness without shifting the threshold. Neither enthalpic stress nor realistic chemical defect fractions can account for experiment. Chemical MRO is therefore the dominant mechanism for rigidity in stoichiometric glasses near the Maxwell point.

The analytical result is exact on tree-like graphs. The pebble-game benchmark confirms the absence of redundant constraints there, but real networks acquire redundancy from short loops~\cite{chubynsky2007}. A first-order expansion of the branching-process argument around the tree-like solution (SM, Sec.~S10) gives $\Delta\langle r\rangle^* \approx -(1-\alpha^*)\,P_{\rm ring}^{\rm eff} \approx -0.57\,P_{\rm ring}^{\rm eff}$: even a modest ring-closure participation of $P_{\rm ring}^{\rm eff}\sim 5\%$, far below the full-rigidification threshold $P_{\rm ring}^*\approx 0.50$, is sufficient to shift $\langle r\rangle^*$ from the mean-field value $2.431$ to the experimental IP lower edge $\approx 2.35$--$2.40$~\cite{jacobs1995,sartbaeva2007,micoulaut2019}. The same argument extends to $\{2,3\}$ networks such as As$_2$Se$_3$, so it is not tied to any specific chemistry. The Maxwell point $\langle r\rangle_c = 2.4$ marks the geometric onset of rigidity percolation in mean-field theory, but chemically mediated medium-range order---not connectivity or geometric topology alone---is required to stabilize rigidity in real stoichiometric glasses. By isolating topology from chemistry, the configuration model provides a zero-MRO baseline that turns stoichiometric rigidity from a phenomenological rule into a quantifiable prediction across material families.

\begin{acknowledgments}
This work was supported by the National High-Level Overseas Talent Program (KS21400126), the Surface and Interface Synthetic Chemistry project (ZXP2025057), the Jiangsu Distinguished Professorship Fund (SR21400225), and the Research Start-up Fund (NH21400525).
\end{acknowledgments}

\noindent\textbf{Data Availability Statement.---} The data that support the findings of this article are openly available in Ref.~\cite{riditiydata2026}.

\clearpage
\onecolumngrid
\begin{center}
  {\Large\bfseries Supplemental Material}
\end{center}

\setcounter{equation}{0}
\setcounter{figure}{0}
\setcounter{table}{0}
\renewcommand{\theequation}{S\arabic{equation}}
\renewcommand{\thefigure}{S\arabic{figure}}
\renewcommand{\thetable}{S\arabic{table}}

\section{S1. Generating-Function Theory Derivations}

In the main text, we define $\alpha = \langle r\rangle - 2$ for the coordination window $\langle r\rangle \in (2, 3)$. The degree distribution for the configuration model consists of $p(2) = 1-\alpha$ and $p(3) = \alpha$. The standard probability generating function for the degree distribution is:
\begin{equation}
  g_0(x) = \sum_{k} p(k) x^k = (1-\alpha)\,x^2 + \alpha\,x^3.
  \label{eq:s_g0}
\end{equation}
The generating function for the excess degree (the number of remaining edges when arriving at a node via a random edge) is defined as $g_1(x) = g_0'(x)/g_0'(1)$, which yields:
\begin{equation}
  g_1(x) = \frac{2(1-\alpha)\,x + 3\alpha\,x^2}{2+\alpha}.
  \label{eq:s_g1}
\end{equation}

In our site percolation process, only nodes with $k_i \geq 3$ are considered ``locally rigid'' and thus retained. We define $u$ as the probability that traversing a random bond does \emph{not} lead to the Giant Rigid Component (GRC). A random bond fails to connect to the GRC if: (i) the node at the other end is $k=2$ (floppy, occurring with probability proportional to the first term in $g_1(1)$), or (ii) the node is $k=3$ (rigid, occurring with probability proportional to the second term in $g_1(1)$), but its remaining two excess edges also fail to connect to the GRC (probability $u^2$). This yields the self-consistency equation:
\begin{equation}
  u = \frac{2(1-\alpha)}{2+\alpha}\cdot 1 + \frac{3\alpha}{2+\alpha}\cdot u^2.
  \label{eq:s_u_derivation}
\end{equation}
Multiplying by $(2+\alpha)$ and rearranging gives the equivalent quadratic form:
\begin{equation}
  3\alpha\,u^2 - (2+\alpha)\,u + 2(1-\alpha) = 0.
\end{equation}

The non-trivial root is:
\begin{equation}
  u^* = \frac{(2+\alpha) - \sqrt{(2+\alpha)^2 - 24\alpha(1-\alpha)}}{6\alpha} = \frac{2(1-\alpha)}{3\alpha}.
\end{equation}
The second equality follows because the discriminant is a perfect square: $(2+\alpha)^2 - 24\alpha(1-\alpha) = (5\alpha-2)^2$.
A macroscopic GRC exists if and only if $u^* < 1$:
\begin{equation}
  \frac{2(1-\alpha)}{3\alpha} < 1 \implies 2 - 2\alpha < 3\alpha \implies \alpha > \frac{2}{5},
\end{equation}
which corresponds to $\langle r\rangle > 2 + 2/5 = 2.4 = \langle r\rangle_c$, recovering the Maxwell isostatic point.

\subsection{Mean-field universality: Extension to $k \in \{2,4\}$ networks at $\langle r\rangle = 2.4$}

The derivation above uses a $k \in \{2,3\}$ distribution appropriate for As$_2$Se$_3$-type systems. Here we show that an analogous derivation for $k \in \{2,4\}$ networks---the natural model for GeSe$_2$ (Ge: $k=4$, Se: $k=2$; stoichiometric $\langle r\rangle = 8/3 \approx 2.67$, here tuned to $\langle r\rangle = 2.4$ via composition)---yields an identical percolation threshold $\langle r\rangle_c = 2.4$, establishing mean-field universality in the mean-field limit.

Consider a configuration-model network with $N$ nodes drawn from a two-point degree distribution: rigid nodes with $k_A = 4$ (fraction $p$) and floppy nodes with $k_B = 2$ (fraction $1-p$). The mean coordination is $\langle r\rangle = 4p + 2(1-p) = 2 + 2p$. At the Maxwell threshold $\langle r\rangle = 2.4$, this corresponds to $p = 0.2$ (20\% rigid nodes), or equivalently a 1:4 number ratio of type-A to type-B atoms rather than the stoichiometric GeSe$_2$ ratio of 1:2. With $\alpha = \langle r\rangle - 2$ as before, we have $p = \alpha/2$ and $1-p = (2-\alpha)/2$.

Within the Phillips-Thorpe framework, the constraint count per node is $C_k = k/2 + \max(0,\,2k-3)$, yielding $C_2 = 2$ and $C_4 = 7$. The local rigidity criterion $C_k \geq 3$ again reduces to $k \geq 3$: nodes with $k=4$ are locally rigid, while $k=2$ nodes are floppy.

The probability of reaching a node of degree $k$ by following a random edge is $kP(k)/\langle r\rangle$. The self-consistency equation for the probability $u$ that a random bond does not lead to the GRC becomes:
\begin{equation}
  u = \frac{2(1-p)}{\langle r\rangle}\cdot 1 + \frac{4p}{\langle r\rangle}\cdot u^3
    = \frac{2-\alpha}{2+\alpha} + \frac{2\alpha}{2+\alpha}\,u^3.
  \label{eq:s_u24}
\end{equation}
Here the $k=2$ term contributes $1$ (floppy, always fails to connect to the GRC) and the $k=4$ term contributes $u^3$ (rigid, but all three excess edges must independently fail). Multiplying by $(2+\alpha)$ and rearranging gives a cubic equation:
\begin{equation}
  2\alpha\,u^3 - (2+\alpha)\,u + (2-\alpha) = 0.
  \label{eq:s_cubic24}
\end{equation}
Since $u=1$ is always a trivial solution, we factor out $(u-1)$:
\begin{equation}
  (u-1)\!\left(2\alpha\,u^2 + 2\alpha\,u - (2-\alpha)\right) = 0.
\end{equation}
The physical branch is determined by the quadratic factor $2\alpha\,u^2 + 2\alpha\,u - (2-\alpha) = 0$, whose positive root is:
\begin{equation}
  u^* = \frac{-\alpha + \sqrt{\alpha(4-\alpha)}}{2\alpha}.
  \label{eq:s_ustar24}
\end{equation}
The GRC emerges when $u^* < 1$. Setting $u^* = 1$:
\begin{equation}
  \sqrt{\alpha(4-\alpha)} = 3\alpha \implies 4\alpha - \alpha^2 = 9\alpha^2 \implies \alpha_c = \frac{2}{5},
\end{equation}
which yields $\langle r\rangle_c = 2 + \alpha_c = 2.4$, identical to the $k \in \{2,3\}$ result.

This universality means that the mean-field rigidity threshold is always $\langle r\rangle_c = 2.4$, regardless of whether the rigid nodes have $k=3$ or $k=4$. Therefore, the topological-mechanical degeneracy is a property of the connectivity rules, not of any specific chemistry.

\section{S2. GRC Fraction and Solution for the 12.5\% Threshold}

The fractional size of the GRC, $S_\infty$, is given by the probability that a randomly chosen node belongs to the rigid cluster. A node belongs to the GRC if it is locally rigid (probability $\alpha$) and at least one of its $k=3$ bonds connects to the GRC (probability $1 - u^3$). Therefore:
\begin{equation}
  S_\infty = \alpha\!\left[1 - (u^*)^3\right] = \alpha\!\left[1 - \!\left(\frac{2(1-\alpha)}{3\alpha}\right)^{\!3}\right].
  \label{eq:s_sinf}
\end{equation}
To find the threshold where the GRC reaches $12.5\%$ of the system size, we set $S_\infty = 0.125$:
\begin{equation}
  0.125 = \alpha - \frac{8(1-\alpha)^3}{27\alpha^2}.
\end{equation}
Multiplying both sides by $27\alpha^2$ and collecting terms gives
\begin{equation}
  35\alpha^3 - \tfrac{219}{8}\,\alpha^2 + 24\alpha - 8 = 0,
  \label{eq:s_cubic}
\end{equation}
or equivalently $280\alpha^3 - 219\alpha^2 + 192\alpha - 64 = 0$ with integer coefficients. Descartes' rule of signs permits up to three positive roots; numerical solution shows only one real root in the physical interval $\alpha \in (2/5, 1)$: $\alpha^* \approx 0.4283$, corresponding to $\langle r\rangle^*_{\rm MF} = 2 + \alpha^* \approx 2.428$.

Table~\ref{tab:mf} compares the analytical mean-field GRC fraction with numerical simulations across the coordination range.

\begin{table}[hbt!]
\caption{\label{tab:mf} Mean-field GRC fraction $S_\infty$ [Eq.~\protect\eqref{eq:s_sinf}] and simulation result ($N=5000$, 20 trials).}
\begin{ruledtabular}
\begin{tabular}{ccccc}
$\langle r\rangle$ & $\alpha$ & $u^*$ & $S_{\infty}^{\rm MF}$ & $S_\infty^{\rm sim}$ \\
\hline
2.40  & 0.400 & 1.000 & 0.000 & $0.040\pm0.020$ \\
2.42  & 0.420 & 0.921 & 0.092 & $0.100\pm0.032$ \\
2.428 & 0.428 & 0.891 & 0.125 & ---             \\
2.43  & 0.430 & 0.884 & 0.133 & $0.117\pm0.048$ \\
2.45  & 0.450 & 0.815 & 0.207 & $0.190\pm0.040$ \\
2.50  & 0.500 & 0.667 & 0.352 & $0.348\pm0.020$ \\
\end{tabular}
\end{ruledtabular}
\end{table}

\section{S3. Maxwell Constraint Counting: Numerical Self-Consistency}

\begin{figure}[hbt!]
  \includegraphics[width=0.7\columnwidth]{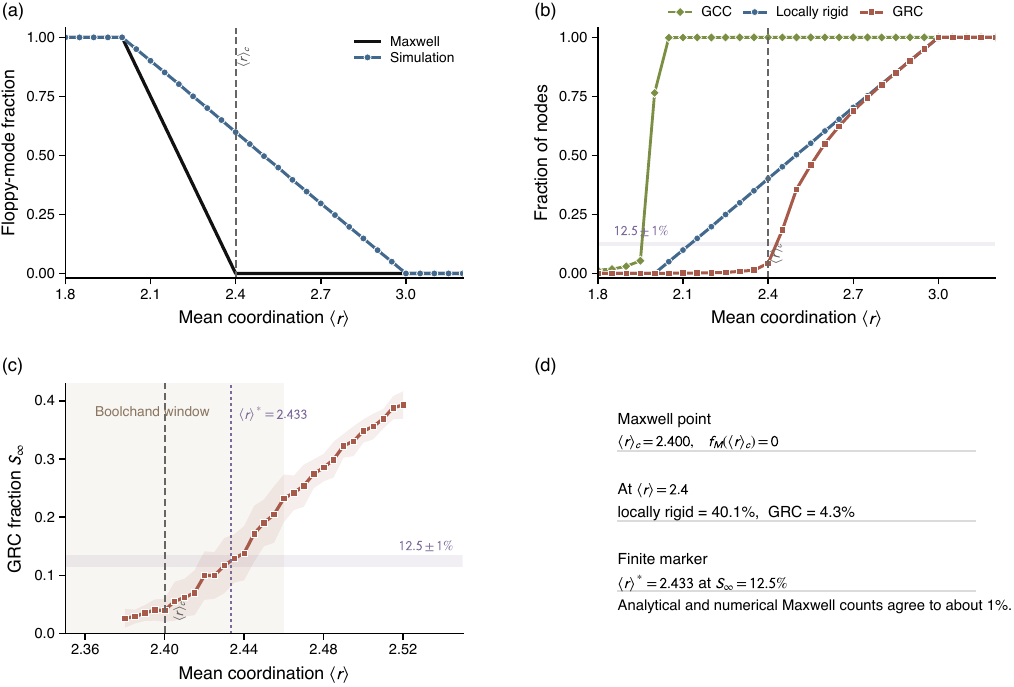}
  \caption{\label{fig:maxwell_valid} Maxwell-count self-consistency. (a)~$f_M$ vs.\ mean coordination $\langle r\rangle$: analytical prediction $f_M = 6-5\langle r\rangle/2$ (solid curve) vs.\ numerical simulation (symbols, $N = 4000$, 10 trials). (b)~Connectivity and rigidity measures on the same coarse sweep. (c)~Fine-resolution view of the rigid backbone near $\langle r\rangle_c = 2.4$, showing the $12.5 \pm 1\%$ extraction band. (d)~Compact summary of the benchmark values quoted in the text.}
\end{figure}

We checked the local criterion $k_i \geq 3$ against the Maxwell count $f_M(\langle r\rangle) = 6 - 5\langle r\rangle/2$~\cite{thorpe1983}. Figure~\ref{fig:maxwell_valid} shows that the agreement is within a few percent across $\langle r\rangle \in [2.0, 3.0]$. Panel (c) also shows the $12.5 \pm 1\%$ extraction band.

Table~\ref{tab:maxwell_check} lists representative values. Then Sec.~S7 addresses non-local redundancy separately.

\begin{table}[hbt!]
\caption{\label{tab:maxwell_check} Analytical vs.\ simulated Maxwell count at selected coordination values ($N = 4000$, 10 trials).}
\begin{ruledtabular}
\begin{tabular}{cccc}
$\langle r\rangle$ & $f^{\rm analytic}$ & $f^{\rm sim}$ & deviation \\
\hline
2.20 & 0.500 & $0.502 \pm 0.004$ & $+0.3\%$ \\
2.40 & 0.000 & $0.003 \pm 0.001$ & $---$ \\
2.50 & $-0.250$ & $-0.248 \pm 0.003$ & $+0.8\%$ \\
2.80 & $-1.000$ & $-0.994 \pm 0.002$ & $+0.6\%$ \\
\end{tabular}
\end{ruledtabular}
\end{table}

\section{S4. ER Percolation Control Experiment}

The main-text marker $S_\infty = 12.5\%$ is already fixed analytically by the rigidity solution in Sec.~S2. The present ER control is therefore interpretive rather than foundational: it shows that the same numerical level also coincides with the classical site-percolation threshold on a locally tree-like graph with mean degree $\langle k\rangle = 8$. For site percolation on an Erd\H{o}s--R\'enyi (ER) graph, the critical occupation probability is $p_c = 1/\langle k\rangle$~\cite{newman2001random}; hence $\langle k\rangle = 8$ gives $p_c = 12.5\%$.

\begin{figure}[hbt!]
  \includegraphics[width=0.7\columnwidth]{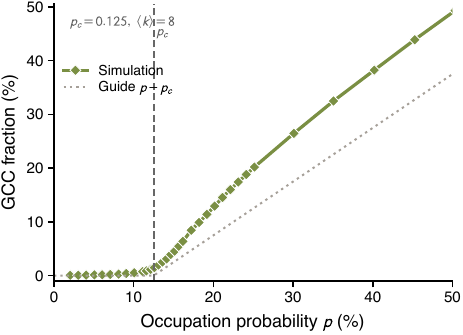}
  \caption{\label{fig:er_control} GCC fraction vs.\ occupation probability $p$ for site percolation on an ER graph with $\langle k\rangle = 8$ ($N = 5000$, 20 trials). The vertical dashed line marks $p_c = 12.5\% = 0.125$.}
\end{figure}

The ER threshold $p_c = 1/\langle k\rangle$ and the rigidity marker $S_\infty = 12.5\%$ both follow from the same branching-process argument in the tree-like limit~\cite{newman2001random,molloy1995}. Therefore, the ER value is a useful sanity check, but the rigidity marker is derived directly from Eq.~(\ref{eq:s_sinf}).

\section{S5. Raw Data for Finite-Size Scaling}

We extrapolated the $12.5\%$ marker to the thermodynamic limit by measuring $\langle r\rangle^*(N)$ for seven system sizes. For each $N$, we generated 40 independent networks and located the crossing by linear interpolation of the mean $S_\infty$ curve to the $12.5\%$ level. In addition, we used the same $12.5 \pm 1\%$ extraction band shown in Fig.~\ref{fig:maxwell_valid}(c). The analytical branch from Sec.~S2 makes the numerical robustness of this choice explicit: shifting the target from $12.5\%$ to $11.5\%$ or $13.5\%$ moves the corresponding mean-field crossing only from $\langle r\rangle = 2.4279$ to $2.4255$ or $2.4304$, i.e.\ by just $\pm 0.0025$ around the central value. The raw finite-size crossing data entering the extrapolation are listed in Table~\ref{tab:fss}.

We fit the data to the power-law scaling relation $\langle r\rangle^*(N) = \langle r\rangle^*_\infty + A\,N^{-1/\nu}$ using all seven system sizes $N \in [500, 32\,000]$. The fit parameters are $\langle r\rangle^*_\infty = 2.431 \pm 0.002$, $A = -1.00 \pm 1.79$, and $\nu = 1.47 \pm 0.43$. Extending to $N = 16\,000$ and $N = 32\,000$ reduces the uncertainty on the extrapolated limit by a factor of four relative to the $N \leq 8\,000$ estimate. The exponent $\nu$ simply tracks how the crossing coordinate approaches its thermodynamic limit; it is \emph{not} the correlation-length critical exponent. In addition, the central result $\langle r\rangle^*_\infty = 2.431 \pm 0.002$ is insensitive to $\nu$: varying it across the full range $(1.0, 1.9)$ changes the extrapolation by less than $0.001$. Uncertainties are estimated via bootstrap resampling ($B=500$ resamples).

\begin{table}[hbt!]
\caption{\label{tab:fss} Extracted $\langle r\rangle$ values at which the GRC fraction reaches $12.5\%$ for various system sizes $N$. The $12.5 \pm 1\%$ extraction window changes the mean-field crossing by only $\pm 0.0025$. Uncertainties are estimated via bootstrap resampling ($B=500$).}
\begin{ruledtabular}
\begin{tabular}{rcc}
$N$ & $\langle r\rangle^*(N)$ & $\pm$ error \\
\hline
500  & 2.4149 & 0.0022 \\
1000 & 2.4194 & 0.0034 \\
2000 & 2.4289 & 0.0020 \\
4000 & 2.4331 & 0.0020 \\
8000 & 2.4306 & 0.0016 \\
16000 & 2.4293 & 0.0012 \\
32000 & 2.4302 & 0.0007 \\
\end{tabular}
\end{ruledtabular}
\end{table}

\section{S6. Topological Structure of the Rigid Backbone via Persistent Homology}

We also applied persistent homology~\cite{hiraoka2016} to the rigid subgraph ($k_i \geq 3$ nodes and their bonds). The Betti numbers $\beta_0$ (connected components) and $\beta_1$ (independent cycles) visualize how the backbone reorganizes across the transition. However, they do not enter the main-text conclusions.

\begin{figure}[hbt!]
  \includegraphics[width=0.7\columnwidth]{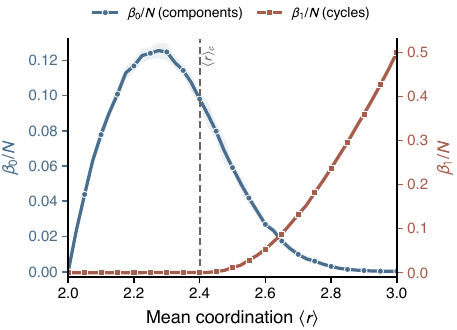}
  \caption{\label{fig:tda} Betti numbers $\beta_0$ (connected components, left axis) and $\beta_1$ (independent cycles, right axis) of the rigid subgraph ($k_i \geq 3$ nodes and their bonds) as a function of mean coordination $\langle r\rangle$ ($N = 3000$, 10 trials). $\beta_0/N$ peaks near $\langle r\rangle \approx 2.3$ (maximum fragmentation) then declines as rigid clusters merge; $\beta_1/N$ rises from zero near $\langle r\rangle \approx 2.4$ as the first macroscopic cycles form.}
\end{figure}

As $\langle r\rangle$ increases, $\beta_0$ first peaks and then falls as fragments merge into a single backbone. At the same time, $\beta_1$ rises from zero near $\langle r\rangle \approx 2.4$ as the first macroscopic cycles form.

\section{S7. Constraint Independence: Laman Pebble Game}

We also checked whether the local criterion $k_i \geq 3$ matches the global constraint count by running the $(2,3)$-Laman pebble game~\cite{jacobs1995,jacobs1996} on the same graphs. The pebble game classifies each edge as \emph{independent} or \emph{redundant}~\cite{laman1970}. Although the $(2,3)$-Laman game is strictly defined for 2D bar-and-joint frameworks, it provides a combinatorial lower bound on redundancy for any graph. Because the configuration model has no spatial embedding and the Phillips-Thorpe constraint count reduces to the same connectivity condition ($k_i \geq 3$), the zero-redundancy result is expected and serves as a consistency check.

\begin{figure}[hbt!]
  \includegraphics[width=0.7\columnwidth]{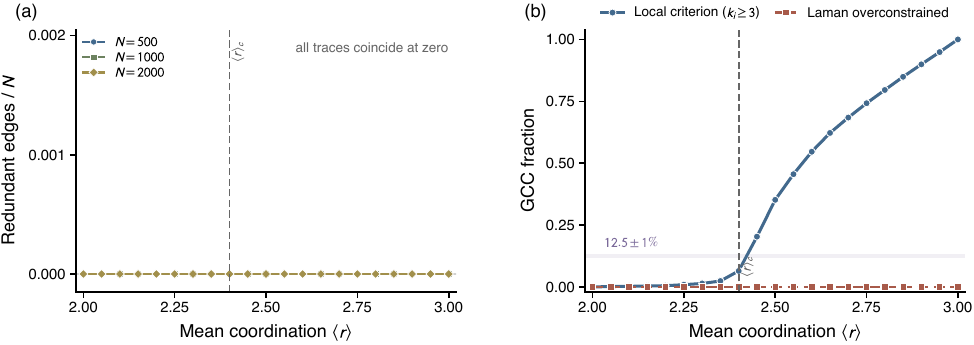}
  \caption{\label{fig:pebble} Laman pebble-game benchmark on configuration-model graphs. (a)~Redundant edge fraction for $N \in \{500, 1000, 2000\}$ (20 trials). The pebble game finds zero redundant edges across the full coordination range, confirming that all constraints are globally independent. (b)~Comparison of the GCC defined by the local criterion ($k_i \geq 3$) with the pebble-game overconstrained region ($N = 2000$). Because no redundancies exist, the overconstrained GCC is identically zero, while the local-criterion GCC tracks the rigid backbone.}
\end{figure}

Figure~\ref{fig:pebble} shows that \emph{every} edge is independent: zero redundancies across all $N$ and $\langle r\rangle$ tested. Therefore, the local criterion $k_i \geq 3$ is consistent with a redundancy-free backbone. This is expected for tree-like graphs, because short loops---the main source of redundancy in embedded networks~\cite{chubynsky2007}---are rare.

\section{S8. External Ring/Linking Analysis Pipeline}

The SiO$_2$ and DNA comparisons in the main text were analyzed with the same ring/linking engine, which is adapted from the procedure of Neophytou et al.~\cite{neophytou2024}.

The SiO$_2$ analysis reports chemical rings and tests for linked-ring pairs. The result is that rings are present but links are absent. For DNA networks, the same linking test is combined with the Fiedler eigenvalue analysis. Therefore, using one engine for both systems ensures a consistent comparison.

\section{S9. Mean-Field Equivalence of Ring Closure and $P_{\rm ring}$ Promotion}

The main-text Test 4 replaces explicit ring-closure motifs---which cannot exist in a locally tree-like graph---with a node-level rigidification: a fraction $P_{\rm ring}$ of $k=2$ nodes is added to the rigid set. At the mean-field level, this is the exact image of ring closure, not a looser proxy. Consider a $k=2$ (B) node $i$ in the tree-like baseline. Its two incident bonds contribute $k/2 = 1$ bond-stretching constraint and $\max(0, 2k-3) = 1$ angular constraint, giving $n_c(i) = 2$. Now close a primitive ring through $i$ by adding one extra bond: the node becomes $k=3$, and its constraint count jumps to $n_c(i) = 3/2 + 3 = 4.5$, identical to a primary rigid ($k=3$) node. Promoting $i$ to the rigid set in the Phillips-Thorpe count is therefore mathematically equivalent to inserting a single ring-closing edge at $i$, modulo the bond-stretching contribution of the extra edge which is shared with the ring partner and does not alter the per-atom floppy-mode count at leading order.

The equivalence is one-to-one per node: $P_{\rm ring}$ as defined in the main text counts the fraction of B nodes whose tree-like environment would be closed by a ring, and each such closure produces exactly the constraint upgrade realized by the rigid-set promotion. As a result, the threshold crossing at $P_{\rm ring}^* \approx 0.50$ can be read directly as the critical ring-participation fraction required to connect rigid clusters through the floppy matrix in a zero-MRO starting configuration.

\section{S10. First-Order Perturbation: Ring-Closure Shift of $\langle r\rangle^*$}

Using the equivalence established in Sec.~S9, the shift of the rigidity threshold under small ring-closure participation $P_{\rm ring}^{\rm eff} \ll 1$ can be obtained analytically. Starting from the $\{2,3\}$ baseline with bare rigid fraction $\alpha = \langle r\rangle - 2$, the effective rigid fraction after ring closure is
\[
\alpha' = \alpha + (1-\alpha)\,P_{\rm ring}^{\rm eff}.
\]
The mean-field Maxwell threshold is at $\alpha' = 2/5$, so the corresponding value of $\langle r\rangle$ satisfies
\[
\langle r\rangle_c(P_{\rm ring}^{\rm eff}) = 2 + \frac{\tfrac{2}{5} - P_{\rm ring}^{\rm eff}}{1 - P_{\rm ring}^{\rm eff}},
\]
which reduces to $\langle r\rangle_c = 2.4$ at $P_{\rm ring}^{\rm eff}=0$. Differentiating at the 12.5\% marker (where $\alpha^* \approx 0.4283$) gives
\[
\left.\frac{d\langle r\rangle^*}{dP_{\rm ring}^{\rm eff}}\right|_{0} = -(1-\alpha^*) \approx -0.57.
\]
To leading order, therefore, $\Delta\langle r\rangle^* \approx -0.57\,P_{\rm ring}^{\rm eff}$. The observed downward shift of approximately $0.03$ from the mean-field marker $\langle r\rangle^* = 2.431$ to the experimental IP lower edge $\langle r\rangle \approx 2.35$--$2.40$ corresponds to $P_{\rm ring}^{\rm eff} \sim 5\%$---far below the full-rigidification threshold $P_{\rm ring}^* \approx 0.50$ but consistent with the short-loop redundancy expected in finite-dimensional covalent networks~\cite{chubynsky2007,jacobs1995}. This partial-participation regime clarifies a point of potential confusion: the experimental IP lies well below $\langle r\rangle^* = 2.431$ not because the mean-field calculation is wrong, but because real glasses realize a finite ring-closure density that perturbs the baseline downward. The mean-field value remains the correct zero-MRO reference against which the experimental shift should be measured.


\begin{thebibliography}{30}%
\makeatletter
\providecommand \@ifxundefined [1]{%
 \@ifx{#1\undefined}
}%
\providecommand \@ifnum [1]{%
 \ifnum #1\expandafter \@firstoftwo
 \else \expandafter \@secondoftwo
 \fi
}%
\providecommand \@ifx [1]{%
 \ifx #1\expandafter \@firstoftwo
 \else \expandafter \@secondoftwo
 \fi
}%
\providecommand \natexlab [1]{#1}%
\providecommand \enquote  [1]{``#1''}%
\providecommand \bibnamefont  [1]{#1}%
\providecommand \bibfnamefont [1]{#1}%
\providecommand \citenamefont [1]{#1}%
\providecommand \href@noop [0]{\@secondoftwo}%
\providecommand \href [0]{\begingroup \@sanitize@url \@href}%
\providecommand \@href[1]{\@@startlink{#1}\@@href}%
\providecommand \@@href[1]{\endgroup#1\@@endlink}%
\providecommand \@sanitize@url [0]{\catcode `\\12\catcode `\$12\catcode
  `\&12\catcode `\#12\catcode `\^12\catcode `\_12\catcode `\%12\relax}%
\providecommand \@@startlink[1]{}%
\providecommand \@@endlink[0]{}%
\providecommand \url  [0]{\begingroup\@sanitize@url \@url }%
\providecommand \@url [1]{\endgroup\@href {#1}{\urlprefix }}%
\providecommand \urlprefix  [0]{URL }%
\providecommand \Eprint [0]{\href }%
\providecommand \doibase [0]{https://doi.org/}%
\providecommand \selectlanguage [0]{\@gobble}%
\providecommand \bibinfo  [0]{\@secondoftwo}%
\providecommand \bibfield  [0]{\@secondoftwo}%
\providecommand \translation [1]{[#1]}%
\providecommand \BibitemOpen [0]{}%
\providecommand \bibitemStop [0]{}%
\providecommand \bibitemNoStop [0]{.\EOS\space}%
\providecommand \EOS [0]{\spacefactor3000\relax}%
\providecommand \BibitemShut  [1]{\csname bibitem#1\endcsname}%
\let\auto@bib@innerbib\@empty
\bibitem [{\citenamefont {Zachariasen}(1932)}]{zachariasen1932}%
  \BibitemOpen
  \bibfield  {author} {\bibinfo {author} {\bibfnamefont {W.~H.}\ \bibnamefont
  {Zachariasen}},\ }\bibfield  {title} {\bibinfo {title} {The atomic
  arrangement in glass},\ }\href {https://doi.org/10.1021/ja01349a006}
  {\bibfield  {journal} {\bibinfo  {journal} {Journal of the American Chemical
  Society}\ }\textbf {\bibinfo {volume} {54}},\ \bibinfo {pages} {3841}
  (\bibinfo {year} {1932})}\BibitemShut {NoStop}%
\bibitem [{\citenamefont {Phillips}(1979)}]{phillips1979}%
  \BibitemOpen
  \bibfield  {author} {\bibinfo {author} {\bibfnamefont {J.~C.}\ \bibnamefont
  {Phillips}},\ }\bibfield  {title} {\bibinfo {title} {Topology of covalent
  non-crystalline solids {I}: Short-range order in chalcogenide alloys},\
  }\href {https://doi.org/10.1016/0022-3093(79)90033-4} {\bibfield  {journal}
  {\bibinfo  {journal} {Journal of Non-Crystalline Solids}\ }\textbf {\bibinfo
  {volume} {34}},\ \bibinfo {pages} {153} (\bibinfo {year} {1979})}\BibitemShut
  {NoStop}%
\bibitem [{\citenamefont {Thorpe}(1983)}]{thorpe1983}%
  \BibitemOpen
  \bibfield  {author} {\bibinfo {author} {\bibfnamefont {M.~F.}\ \bibnamefont
  {Thorpe}},\ }\bibfield  {title} {\bibinfo {title} {Continuous deformations in
  random networks},\ }\href {https://doi.org/10.1016/0022-3093(83)90424-6}
  {\bibfield  {journal} {\bibinfo  {journal} {Journal of Non-Crystalline
  Solids}\ }\textbf {\bibinfo {volume} {57}},\ \bibinfo {pages} {355} (\bibinfo
  {year} {1983})}\BibitemShut {NoStop}%
\bibitem [{\citenamefont {Jacobs}\ and\ \citenamefont
  {Thorpe}(1995)}]{jacobs1995}%
  \BibitemOpen
  \bibfield  {author} {\bibinfo {author} {\bibfnamefont {D.~J.}\ \bibnamefont
  {Jacobs}}\ and\ \bibinfo {author} {\bibfnamefont {M.~F.}\ \bibnamefont
  {Thorpe}},\ }\bibfield  {title} {\bibinfo {title} {Generic rigidity
  percolation: The pebble game},\ }\href
  {https://doi.org/10.1103/PhysRevLett.75.4051} {\bibfield  {journal} {\bibinfo
   {journal} {Physical Review Letters}\ }\textbf {\bibinfo {volume} {75}},\
  \bibinfo {pages} {4051} (\bibinfo {year} {1995})}\BibitemShut {NoStop}%
\bibitem [{\citenamefont {Micoulaut}\ \emph {et~al.}(2022)\citenamefont
  {Micoulaut}, \citenamefont {Pethes}, \citenamefont {J{\'o}v{\'a}ri},
  \citenamefont {Pusztai}, \citenamefont {Krbal}, \citenamefont {W{\'a}gner},
  \citenamefont {Prokop}, \citenamefont {Michalik}, \citenamefont {Ikeda},\
  and\ \citenamefont {Kaban}}]{micoulaut2022}%
  \BibitemOpen
  \bibfield  {author} {\bibinfo {author} {\bibfnamefont {M.}~\bibnamefont
  {Micoulaut}}, \bibinfo {author} {\bibfnamefont {I.}~\bibnamefont {Pethes}},
  \bibinfo {author} {\bibfnamefont {P.}~\bibnamefont {J{\'o}v{\'a}ri}},
  \bibinfo {author} {\bibfnamefont {L.}~\bibnamefont {Pusztai}}, \bibinfo
  {author} {\bibfnamefont {M.}~\bibnamefont {Krbal}}, \bibinfo {author}
  {\bibfnamefont {T.}~\bibnamefont {W{\'a}gner}}, \bibinfo {author}
  {\bibfnamefont {V.}~\bibnamefont {Prokop}}, \bibinfo {author} {\bibfnamefont
  {{\v{S}}.}~\bibnamefont {Michalik}}, \bibinfo {author} {\bibfnamefont
  {K.}~\bibnamefont {Ikeda}},\ and\ \bibinfo {author} {\bibfnamefont
  {I.}~\bibnamefont {Kaban}},\ }\bibfield  {title} {\bibinfo {title}
  {Structural properties of chalcogenide glasses and the isocoordination rule:
  Disentangling effects from chemistry and network topology},\ }\href
  {https://doi.org/10.1103/PhysRevB.106.014206} {\bibfield  {journal} {\bibinfo
   {journal} {Physical Review B}\ }\textbf {\bibinfo {volume} {106}},\ \bibinfo
  {pages} {014206} (\bibinfo {year} {2022})}\BibitemShut {NoStop}%
\bibitem [{\citenamefont {Chubynsky}\ and\ \citenamefont
  {Thorpe}(2007)}]{chubynsky2007}%
  \BibitemOpen
  \bibfield  {author} {\bibinfo {author} {\bibfnamefont {M.~V.}\ \bibnamefont
  {Chubynsky}}\ and\ \bibinfo {author} {\bibfnamefont {M.~F.}\ \bibnamefont
  {Thorpe}},\ }\bibfield  {title} {\bibinfo {title} {Algorithms for
  three-dimensional rigidity analysis and a first-order percolation
  transition},\ }\href {https://doi.org/10.1103/PhysRevE.76.041135} {\bibfield
  {journal} {\bibinfo  {journal} {Physical Review E}\ }\textbf {\bibinfo
  {volume} {76}},\ \bibinfo {pages} {041135} (\bibinfo {year}
  {2007})}\BibitemShut {NoStop}%
\bibitem [{\citenamefont {Micoulaut}(2023)}]{micoulaut2023review}%
  \BibitemOpen
  \bibfield  {author} {\bibinfo {author} {\bibfnamefont {M.}~\bibnamefont
  {Micoulaut}},\ }\bibfield  {title} {\bibinfo {title} {Topological ordering
  during flexible to rigid transitions in disordered networks},\ }\href
  {https://doi.org/10.5802/crphys.128} {\bibfield  {journal} {\bibinfo
  {journal} {Comptes Rendus Physique}\ }\textbf {\bibinfo {volume} {24}},\
  \bibinfo {pages} {133} (\bibinfo {year} {2023})}\BibitemShut {NoStop}%
\bibitem [{\citenamefont {Broedersz}\ and\ \citenamefont
  {MacKintosh}(2014)}]{broedersz2014}%
  \BibitemOpen
  \bibfield  {author} {\bibinfo {author} {\bibfnamefont {C.~P.}\ \bibnamefont
  {Broedersz}}\ and\ \bibinfo {author} {\bibfnamefont {F.~C.}\ \bibnamefont
  {MacKintosh}},\ }\bibfield  {title} {\bibinfo {title} {Modeling semiflexible
  polymer networks},\ }\href {https://doi.org/10.1103/RevModPhys.86.995}
  {\bibfield  {journal} {\bibinfo  {journal} {Reviews of Modern Physics}\
  }\textbf {\bibinfo {volume} {86}},\ \bibinfo {pages} {995} (\bibinfo {year}
  {2014})}\BibitemShut {NoStop}%
\bibitem [{\citenamefont {Zaccone}\ and\ \citenamefont
  {Scossa-Romano}(2019)}]{zaccone2019correlated}%
  \BibitemOpen
  \bibfield  {author} {\bibinfo {author} {\bibfnamefont {A.}~\bibnamefont
  {Zaccone}}\ and\ \bibinfo {author} {\bibfnamefont {E.}~\bibnamefont
  {Scossa-Romano}},\ }\bibfield  {title} {\bibinfo {title} {Correlated rigidity
  percolation and colloidal gels},\ }\href
  {https://doi.org/10.1103/PhysRevLett.123.058001} {\bibfield  {journal}
  {\bibinfo  {journal} {Physical Review Letters}\ }\textbf {\bibinfo {volume}
  {123}},\ \bibinfo {pages} {058001} (\bibinfo {year} {2019})}\BibitemShut
  {NoStop}%
\bibitem [{\citenamefont {Luo}\ \emph {et~al.}(2019)\citenamefont {Luo},
  \citenamefont {Vermant},\ and\ \citenamefont {Ilg}}]{luo2019direct}%
  \BibitemOpen
  \bibfield  {author} {\bibinfo {author} {\bibfnamefont {J.}~\bibnamefont
  {Luo}}, \bibinfo {author} {\bibfnamefont {J.}~\bibnamefont {Vermant}},\ and\
  \bibinfo {author} {\bibfnamefont {P.}~\bibnamefont {Ilg}},\ }\bibfield
  {title} {\bibinfo {title} {Direct link between mechanical stability in gels
  and percolation of isostatic particles},\ }\href
  {https://doi.org/10.1126/sciadv.aav6090} {\bibfield  {journal} {\bibinfo
  {journal} {Science Advances}\ }\textbf {\bibinfo {volume} {5}},\ \bibinfo
  {pages} {eaav6090} (\bibinfo {year} {2019})}\BibitemShut {NoStop}%
\bibitem [{\citenamefont {Nishikawa}\ \emph {et~al.}(2022)\citenamefont
  {Nishikawa}, \citenamefont {Okano}, \citenamefont {Kodera}, \citenamefont
  {Tanabe},\ and\ \citenamefont {Nakata}}]{nishikawa2022rigidity}%
  \BibitemOpen
  \bibfield  {author} {\bibinfo {author} {\bibfnamefont {M.}~\bibnamefont
  {Nishikawa}}, \bibinfo {author} {\bibfnamefont {K.}~\bibnamefont {Okano}},
  \bibinfo {author} {\bibfnamefont {R.}~\bibnamefont {Kodera}}, \bibinfo
  {author} {\bibfnamefont {H.}~\bibnamefont {Tanabe}},\ and\ \bibinfo {author}
  {\bibfnamefont {M.}~\bibnamefont {Nakata}},\ }\bibfield  {title} {\bibinfo
  {title} {Rigidity-dependent formation process of dna supramolecular
  hydrogels},\ }\href {https://doi.org/10.1038/s41427-022-00445-w} {\bibfield
  {journal} {\bibinfo  {journal} {NPG Asia Materials}\ }\textbf {\bibinfo
  {volume} {14}},\ \bibinfo {pages} {94} (\bibinfo {year} {2022})}\BibitemShut
  {NoStop}%
\bibitem [{\citenamefont {Neves}\ \emph {et~al.}(2025)\citenamefont {Neves},
  \citenamefont {Tavares}, \citenamefont {Ara{\'u}jo},\ and\ \citenamefont
  {Dias}}]{dias2025}%
  \BibitemOpen
  \bibfield  {author} {\bibinfo {author} {\bibfnamefont {J.~C.}\ \bibnamefont
  {Neves}}, \bibinfo {author} {\bibfnamefont {J.~M.}\ \bibnamefont {Tavares}},
  \bibinfo {author} {\bibfnamefont {N.~A.~M.}\ \bibnamefont {Ara{\'u}jo}},\
  and\ \bibinfo {author} {\bibfnamefont {C.~S.}\ \bibnamefont {Dias}},\
  }\bibfield  {title} {\bibinfo {title} {Rigid m-percolation in limited-valence
  gels},\ }\href@noop {} {\bibfield  {journal} {\bibinfo  {journal} {arXiv
  preprint arXiv:2504.02474}\ } (\bibinfo {year} {2025})},\ \Eprint
  {https://arxiv.org/abs/2504.02474} {arXiv:2504.02474} \BibitemShut {NoStop}%
\bibitem [{\citenamefont {Neophytou}\ \emph {et~al.}(2024)\citenamefont
  {Neophytou}, \citenamefont {Starr}, \citenamefont {Chakrabarti},\ and\
  \citenamefont {Sciortino}}]{neophytou2024}%
  \BibitemOpen
  \bibfield  {author} {\bibinfo {author} {\bibfnamefont {A.}~\bibnamefont
  {Neophytou}}, \bibinfo {author} {\bibfnamefont {F.~W.}\ \bibnamefont
  {Starr}}, \bibinfo {author} {\bibfnamefont {D.}~\bibnamefont {Chakrabarti}},\
  and\ \bibinfo {author} {\bibfnamefont {F.}~\bibnamefont {Sciortino}},\
  }\bibfield  {title} {\bibinfo {title} {Hierarchy of topological transitions
  in a network liquid},\ }\href {https://doi.org/10.1073/pnas.2406890121}
  {\bibfield  {journal} {\bibinfo  {journal} {Proceedings of the National
  Academy of Sciences}\ }\textbf {\bibinfo {volume} {121}},\ \bibinfo {pages}
  {e2406890121} (\bibinfo {year} {2024})}\BibitemShut {NoStop}%
\bibitem [{\citenamefont {Micoulaut}\ and\ \citenamefont
  {Phillips}(2003)}]{micoulaut2003}%
  \BibitemOpen
  \bibfield  {author} {\bibinfo {author} {\bibfnamefont {M.}~\bibnamefont
  {Micoulaut}}\ and\ \bibinfo {author} {\bibfnamefont {J.~C.}\ \bibnamefont
  {Phillips}},\ }\bibfield  {title} {\bibinfo {title} {Rings and rigidity
  transitions in network glasses},\ }\href
  {https://doi.org/10.1103/PhysRevB.67.104204} {\bibfield  {journal} {\bibinfo
  {journal} {Physical Review B}\ }\textbf {\bibinfo {volume} {67}},\ \bibinfo
  {pages} {104204} (\bibinfo {year} {2003})}\BibitemShut {NoStop}%
\bibitem [{\citenamefont {Molloy}\ and\ \citenamefont
  {Reed}(1995)}]{molloy1995}%
  \BibitemOpen
  \bibfield  {author} {\bibinfo {author} {\bibfnamefont {M.}~\bibnamefont
  {Molloy}}\ and\ \bibinfo {author} {\bibfnamefont {B.}~\bibnamefont {Reed}},\
  }\bibfield  {title} {\bibinfo {title} {A critical point for random graphs
  with a given degree sequence},\ }\href
  {https://doi.org/10.1002/rsa.3240060204} {\bibfield  {journal} {\bibinfo
  {journal} {Random Structures \& Algorithms}\ }\textbf {\bibinfo {volume}
  {6}},\ \bibinfo {pages} {161} (\bibinfo {year} {1995})}\BibitemShut {NoStop}%
\bibitem [{\citenamefont {Sartbaeva}\ \emph {et~al.}(2007)\citenamefont
  {Sartbaeva}, \citenamefont {Wells}, \citenamefont {Thorpe}, \citenamefont
  {Bychkov},\ and\ \citenamefont {Benmore}}]{sartbaeva2007}%
  \BibitemOpen
  \bibfield  {author} {\bibinfo {author} {\bibfnamefont {A.}~\bibnamefont
  {Sartbaeva}}, \bibinfo {author} {\bibfnamefont {S.~A.}\ \bibnamefont
  {Wells}}, \bibinfo {author} {\bibfnamefont {M.~F.}\ \bibnamefont {Thorpe}},
  \bibinfo {author} {\bibfnamefont {E.}~\bibnamefont {Bychkov}},\ and\ \bibinfo
  {author} {\bibfnamefont {C.~J.}\ \bibnamefont {Benmore}},\ }\bibfield
  {title} {\bibinfo {title} {Local structural variability and the intermediate
  phase window in \texorpdfstring{Ge$_x$Se$_{1-x}$}{GexSe1-x} glasses},\ }\href
  {https://doi.org/10.1103/PhysRevB.75.224204} {\bibfield  {journal} {\bibinfo
  {journal} {Physical Review B}\ }\textbf {\bibinfo {volume} {75}},\ \bibinfo
  {pages} {224204} (\bibinfo {year} {2007})}\BibitemShut {NoStop}%
\bibitem [{\citenamefont {Zeidler}\ \emph {et~al.}(2017)\citenamefont
  {Zeidler}, \citenamefont {Salmon},\ and\ \citenamefont
  {Hannon}}]{zeidler2017}%
  \BibitemOpen
  \bibfield  {author} {\bibinfo {author} {\bibfnamefont {A.}~\bibnamefont
  {Zeidler}}, \bibinfo {author} {\bibfnamefont {P.~S.}\ \bibnamefont
  {Salmon}},\ and\ \bibinfo {author} {\bibfnamefont {A.~C.}\ \bibnamefont
  {Hannon}},\ }\bibfield  {title} {\bibinfo {title} {Topological ordering and
  viscosity in the glass-forming \texorpdfstring{Ge$_x$Se$_{1-x}$}{GexSe1-x}
  system: An intermediate phase interpretation},\ }\href
  {https://doi.org/10.3389/fmats.2017.00032} {\bibfield  {journal} {\bibinfo
  {journal} {Frontiers in Materials}\ }\textbf {\bibinfo {volume} {4}},\
  \bibinfo {pages} {32} (\bibinfo {year} {2017})}\BibitemShut {NoStop}%
\bibitem [{\citenamefont {Sharma}\ \emph {et~al.}(1981)\citenamefont {Sharma},
  \citenamefont {Mammone},\ and\ \citenamefont {Nicol}}]{sharma1981}%
  \BibitemOpen
  \bibfield  {author} {\bibinfo {author} {\bibfnamefont {S.~K.}\ \bibnamefont
  {Sharma}}, \bibinfo {author} {\bibfnamefont {J.~F.}\ \bibnamefont
  {Mammone}},\ and\ \bibinfo {author} {\bibfnamefont {M.~F.}\ \bibnamefont
  {Nicol}},\ }\bibfield  {title} {\bibinfo {title} {Raman investigation of ring
  configurations in vitreous silica},\ }\href
  {https://doi.org/10.1038/292140a0} {\bibfield  {journal} {\bibinfo  {journal}
  {Nature}\ }\textbf {\bibinfo {volume} {292}},\ \bibinfo {pages} {140}
  (\bibinfo {year} {1981})}\BibitemShut {NoStop}%
\bibitem [{\citenamefont {Micoulaut}\ and\ \citenamefont
  {Bauchy}(2019)}]{micoulaut2019}%
  \BibitemOpen
  \bibfield  {author} {\bibinfo {author} {\bibfnamefont {M.}~\bibnamefont
  {Micoulaut}}\ and\ \bibinfo {author} {\bibfnamefont {M.}~\bibnamefont
  {Bauchy}},\ }\bibfield  {title} {\bibinfo {title} {Statistical mechanical
  model of the self-organized intermediate phase in glass-forming systems with
  adaptable network topologies},\ }\href
  {https://doi.org/10.3389/fmats.2019.00011} {\bibfield  {journal} {\bibinfo
  {journal} {Frontiers in Materials}\ }\textbf {\bibinfo {volume} {6}},\
  \bibinfo {pages} {11} (\bibinfo {year} {2019})}\BibitemShut {NoStop}%
\bibitem [{\citenamefont {Gjersing}\ \emph {et~al.}(2010)\citenamefont
  {Gjersing}, \citenamefont {Sen},\ and\ \citenamefont
  {Aitken}}]{gjersing2010}%
  \BibitemOpen
  \bibfield  {author} {\bibinfo {author} {\bibfnamefont {E.~L.}\ \bibnamefont
  {Gjersing}}, \bibinfo {author} {\bibfnamefont {S.}~\bibnamefont {Sen}},\ and\
  \bibinfo {author} {\bibfnamefont {B.~G.}\ \bibnamefont {Aitken}},\ }\bibfield
   {title} {\bibinfo {title} {Structure, connectivity, and configurational
  entropy of \texorpdfstring{Ge$_x$Se$_{100-x}$}{GexSe100-x} glasses: Results
  from \texorpdfstring{$^{77}$Se}{77Se} mas nmr spectroscopy},\ }\href
  {https://doi.org/10.1021/jp908737e} {\bibfield  {journal} {\bibinfo
  {journal} {Journal of Physical Chemistry C}\ }\textbf {\bibinfo {volume}
  {114}},\ \bibinfo {pages} {519} (\bibinfo {year} {2010})}\BibitemShut
  {NoStop}%
\bibitem [{\citenamefont {Tavanti}\ \emph {et~al.}(2020)\citenamefont
  {Tavanti}, \citenamefont {Dianat}, \citenamefont {Catellani},\ and\
  \citenamefont {Calzolari}}]{tavanti2020}%
  \BibitemOpen
  \bibfield  {author} {\bibinfo {author} {\bibfnamefont {F.}~\bibnamefont
  {Tavanti}}, \bibinfo {author} {\bibfnamefont {B.}~\bibnamefont {Dianat}},
  \bibinfo {author} {\bibfnamefont {A.}~\bibnamefont {Catellani}},\ and\
  \bibinfo {author} {\bibfnamefont {A.}~\bibnamefont {Calzolari}},\ }\bibfield
  {title} {\bibinfo {title} {Hierarchical short- and medium-range order
  structures in amorphous gexse1-x for selectors applications},\ }\href
  {https://doi.org/10.1021/acsaelm.0c00581} {\bibfield  {journal} {\bibinfo
  {journal} {ACS Applied Electronic Materials}\ }\textbf {\bibinfo {volume}
  {2}},\ \bibinfo {pages} {2961} (\bibinfo {year} {2020})}\BibitemShut
  {NoStop}%
\bibitem [{\citenamefont {Tsiulyanu}\ \emph {et~al.}(2022)\citenamefont
  {Tsiulyanu}, \citenamefont {Kozyukhin},\ and\ \citenamefont
  {Ciobanu}}]{tsiulyanu2022}%
  \BibitemOpen
  \bibfield  {author} {\bibinfo {author} {\bibfnamefont {D.}~\bibnamefont
  {Tsiulyanu}}, \bibinfo {author} {\bibfnamefont {S.~A.}\ \bibnamefont
  {Kozyukhin}},\ and\ \bibinfo {author} {\bibfnamefont {M.}~\bibnamefont
  {Ciobanu}},\ }\bibfield  {title} {\bibinfo {title} {Middle range order and
  elastic properties of non-stoichiometric chalcogenide glasses in the
  ass3-ges4 system},\ }\href {https://doi.org/10.1016/j.jnoncrysol.2021.121207}
  {\bibfield  {journal} {\bibinfo  {journal} {Journal of Non-Crystalline
  Solids}\ }\textbf {\bibinfo {volume} {575}},\ \bibinfo {pages} {121207}
  (\bibinfo {year} {2022})}\BibitemShut {NoStop}%
\bibitem [{\citenamefont {Sidebottom}\ and\ \citenamefont
  {Olabode}(2025)}]{sidebottom2025}%
  \BibitemOpen
  \bibfield  {author} {\bibinfo {author} {\bibfnamefont {D.~L.}\ \bibnamefont
  {Sidebottom}}\ and\ \bibinfo {author} {\bibfnamefont {D.}~\bibnamefont
  {Olabode}},\ }\bibfield  {title} {\bibinfo {title} {Network topology uniquely
  determines glass fragility},\ }\href {https://doi.org/10.1103/2m4g-w4yk}
  {\bibfield  {journal} {\bibinfo  {journal} {Physical Review Materials}\
  }\textbf {\bibinfo {volume} {9}},\ \bibinfo {pages} {105601} (\bibinfo {year}
  {2025})}\BibitemShut {NoStop}%
\bibitem [{\citenamefont {Jain}\ \emph {et~al.}(2013)\citenamefont {Jain},
  \citenamefont {Ong}, \citenamefont {Hautier}, \citenamefont {Chen},
  \citenamefont {Richards}, \citenamefont {Dacek}, \citenamefont {Cholia},
  \citenamefont {Gunter}, \citenamefont {Skinner}, \citenamefont {Ceder},\ and\
  \citenamefont {Persson}}]{materialsproject}%
  \BibitemOpen
  \bibfield  {author} {\bibinfo {author} {\bibfnamefont {A.}~\bibnamefont
  {Jain}}, \bibinfo {author} {\bibfnamefont {S.~P.}\ \bibnamefont {Ong}},
  \bibinfo {author} {\bibfnamefont {G.}~\bibnamefont {Hautier}}, \bibinfo
  {author} {\bibfnamefont {W.}~\bibnamefont {Chen}}, \bibinfo {author}
  {\bibfnamefont {W.~D.}\ \bibnamefont {Richards}}, \bibinfo {author}
  {\bibfnamefont {S.}~\bibnamefont {Dacek}}, \bibinfo {author} {\bibfnamefont
  {S.}~\bibnamefont {Cholia}}, \bibinfo {author} {\bibfnamefont
  {D.}~\bibnamefont {Gunter}}, \bibinfo {author} {\bibfnamefont
  {D.}~\bibnamefont {Skinner}}, \bibinfo {author} {\bibfnamefont
  {G.}~\bibnamefont {Ceder}},\ and\ \bibinfo {author} {\bibfnamefont {K.~A.}\
  \bibnamefont {Persson}},\ }\bibfield  {title} {\bibinfo {title} {Commentary:
  The materials project: A materials genome approach to accelerating materials
  innovation},\ }\href {https://doi.org/10.1063/1.4812323} {\bibfield
  {journal} {\bibinfo  {journal} {APL Materials}\ }\textbf {\bibinfo {volume}
  {1}},\ \bibinfo {pages} {011002} (\bibinfo {year} {2013})}\BibitemShut
  {NoStop}%
\bibitem [{\citenamefont {Boolchand}\ \emph {et~al.}(2001)\citenamefont
  {Boolchand}, \citenamefont {Georgiev},\ and\ \citenamefont
  {Goodman}}]{boolchand2001}%
  \BibitemOpen
  \bibfield  {author} {\bibinfo {author} {\bibfnamefont {P.}~\bibnamefont
  {Boolchand}}, \bibinfo {author} {\bibfnamefont {D.~G.}\ \bibnamefont
  {Georgiev}},\ and\ \bibinfo {author} {\bibfnamefont {B.}~\bibnamefont
  {Goodman}},\ }\bibfield  {title} {\bibinfo {title} {Discovery of the
  intermediate phase in chalcogenide glasses},\ }\href
  {https://doi.org/10.1016/S0022-3093(01)00867-5} {\bibfield  {journal}
  {\bibinfo  {journal} {Journal of Non-Crystalline Solids}\ }\textbf {\bibinfo
  {volume} {293--295}},\ \bibinfo {pages} {348} (\bibinfo {year}
  {2001})}\BibitemShut {NoStop}%
\bibitem [{\citenamefont {Boolchand}\ \emph {et~al.}(2005)\citenamefont
  {Boolchand}, \citenamefont {Lucovsky}, \citenamefont {Phillips},\ and\
  \citenamefont {Thorpe}}]{boolchand2005}%
  \BibitemOpen
  \bibfield  {author} {\bibinfo {author} {\bibfnamefont {P.}~\bibnamefont
  {Boolchand}}, \bibinfo {author} {\bibfnamefont {G.}~\bibnamefont {Lucovsky}},
  \bibinfo {author} {\bibfnamefont {J.~C.}\ \bibnamefont {Phillips}},\ and\
  \bibinfo {author} {\bibfnamefont {M.~F.}\ \bibnamefont {Thorpe}},\ }\bibfield
   {title} {\bibinfo {title} {Self-organization and the physics of glassy
  networks},\ }\href {https://doi.org/10.1080/14786430500256425} {\bibfield
  {journal} {\bibinfo  {journal} {Philosophical Magazine}\ }\textbf {\bibinfo
  {volume} {85}},\ \bibinfo {pages} {3823} (\bibinfo {year}
  {2005})}\BibitemShut {NoStop}%
\bibitem [{\citenamefont {Thorpe}\ \emph {et~al.}(2000)\citenamefont {Thorpe},
  \citenamefont {Jacobs}, \citenamefont {Chubynsky},\ and\ \citenamefont
  {Phillips}}]{thorpe2000}%
  \BibitemOpen
  \bibfield  {author} {\bibinfo {author} {\bibfnamefont {M.~F.}\ \bibnamefont
  {Thorpe}}, \bibinfo {author} {\bibfnamefont {D.~J.}\ \bibnamefont {Jacobs}},
  \bibinfo {author} {\bibfnamefont {M.~V.}\ \bibnamefont {Chubynsky}},\ and\
  \bibinfo {author} {\bibfnamefont {J.~C.}\ \bibnamefont {Phillips}},\
  }\bibfield  {title} {\bibinfo {title} {Self-organization in network
  glasses},\ }\href {https://doi.org/10.1016/S0022-3093(99)00856-X} {\bibfield
  {journal} {\bibinfo  {journal} {Journal of Non-Crystalline Solids}\ }\textbf
  {\bibinfo {volume} {266--269}},\ \bibinfo {pages} {859} (\bibinfo {year}
  {2000})}\BibitemShut {NoStop}%
\bibitem [{\citenamefont {Conrad}\ \emph {et~al.}(2019)\citenamefont {Conrad},
  \citenamefont {Kennedy}, \citenamefont {Fygenson},\ and\ \citenamefont
  {Saleh}}]{conrad2019}%
  \BibitemOpen
  \bibfield  {author} {\bibinfo {author} {\bibfnamefont {N.}~\bibnamefont
  {Conrad}}, \bibinfo {author} {\bibfnamefont {T.}~\bibnamefont {Kennedy}},
  \bibinfo {author} {\bibfnamefont {D.~K.}\ \bibnamefont {Fygenson}},\ and\
  \bibinfo {author} {\bibfnamefont {O.~A.}\ \bibnamefont {Saleh}},\ }\bibfield
  {title} {\bibinfo {title} {Increasing valence pushes dna nanostar networks to
  the isostatic point},\ }\href {https://doi.org/10.1073/pnas.1819683116}
  {\bibfield  {journal} {\bibinfo  {journal} {Proceedings of the National
  Academy of Sciences of the United States of America}\ }\textbf {\bibinfo
  {volume} {116}},\ \bibinfo {pages} {7238} (\bibinfo {year}
  {2019})}\BibitemShut {NoStop}%
\bibitem [{\citenamefont {Palombo}\ \emph {et~al.}(2025)\citenamefont
  {Palombo}, \citenamefont {Weir}, \citenamefont {Michieletto},\ and\
  \citenamefont {Guti{\'e}rrez~Fosado}}]{palombo2025}%
  \BibitemOpen
  \bibfield  {author} {\bibinfo {author} {\bibfnamefont {G.}~\bibnamefont
  {Palombo}}, \bibinfo {author} {\bibfnamefont {S.}~\bibnamefont {Weir}},
  \bibinfo {author} {\bibfnamefont {D.}~\bibnamefont {Michieletto}},\ and\
  \bibinfo {author} {\bibfnamefont {Y.~A.}\ \bibnamefont
  {Guti{\'e}rrez~Fosado}},\ }\bibfield  {title} {\bibinfo {title} {Topological
  linking determines elasticity in limited valence networks},\ }\href
  {https://doi.org/10.1038/s41563-024-02091-9} {\bibfield  {journal} {\bibinfo
  {journal} {Nature Materials}\ }\textbf {\bibinfo {volume} {24}},\ \bibinfo
  {pages} {454} (\bibinfo {year} {2025})}\BibitemShut {NoStop}%
\bibitem [{\citenamefont {Liu}(2026)}]{riditiydata2026}%
  \BibitemOpen
  \bibfield  {author} {\bibinfo {author} {\bibfnamefont {K.}~\bibnamefont
  {Liu}},\ }\href@noop {} {\bibinfo {title} {Numerical data and analysis code
  for ``chemical medium-range order enables stoichiometric rigidity''}},\
  \bibinfo {howpublished}
  {\url{https://github.com/KejunLiuGitHub/riditiy-data}} (\bibinfo {year}
  {2026}),\ \bibinfo {note} {accessed: 2026-04-25}\BibitemShut {NoStop}%
\end{thebibliography}

\begin{thebibliography}{9}%
\makeatletter
\providecommand \@ifxundefined [1]{%
 \@ifx{#1\undefined}
}%
\providecommand \@ifnum [1]{%
 \ifnum #1\expandafter \@firstoftwo
 \else \expandafter \@secondoftwo
 \fi
}%
\providecommand \@ifx [1]{%
 \ifx #1\expandafter \@firstoftwo
 \else \expandafter \@secondoftwo
 \fi
}%
\providecommand \natexlab [1]{#1}%
\providecommand \enquote  [1]{``#1''}%
\providecommand \bibnamefont  [1]{#1}%
\providecommand \bibfnamefont [1]{#1}%
\providecommand \citenamefont [1]{#1}%
\providecommand \href@noop [0]{\@secondoftwo}%
\providecommand \href [0]{\begingroup \@sanitize@url \@href}%
\providecommand \@href[1]{\@@startlink{#1}\@@href}%
\providecommand \@@href[1]{\endgroup#1\@@endlink}%
\providecommand \@sanitize@url [0]{\catcode `\\12\catcode `\$12\catcode
  `\&12\catcode `\#12\catcode `\^12\catcode `\_12\catcode `\%12\relax}%
\providecommand \@@startlink[1]{}%
\providecommand \@@endlink[0]{}%
\providecommand \url  [0]{\begingroup\@sanitize@url \@url }%
\providecommand \@url [1]{\endgroup\@href {#1}{\urlprefix }}%
\providecommand \urlprefix  [0]{URL }%
\providecommand \Eprint [0]{\href }%
\providecommand \doibase [0]{https://doi.org/}%
\providecommand \selectlanguage [0]{\@gobble}%
\providecommand \bibinfo  [0]{\@secondoftwo}%
\providecommand \bibfield  [0]{\@secondoftwo}%
\providecommand \translation [1]{[#1]}%
\providecommand \BibitemOpen [0]{}%
\providecommand \bibitemStop [0]{}%
\providecommand \bibitemNoStop [0]{.\EOS\space}%
\providecommand \EOS [0]{\spacefactor3000\relax}%
\providecommand \BibitemShut  [1]{\csname bibitem#1\endcsname}%
\let\auto@bib@innerbib\@empty
\bibitem [{\citenamefont {Thorpe}(1983)}]{thorpe1983}%
  \BibitemOpen
  \bibfield  {author} {\bibinfo {author} {\bibfnamefont {M.~F.}\ \bibnamefont
  {Thorpe}},\ }\bibfield  {title} {\bibinfo {title} {Continuous deformations in
  random networks},\ }\href {https://doi.org/10.1016/0022-3093(83)90424-6}
  {\bibfield  {journal} {\bibinfo  {journal} {Journal of Non-Crystalline
  Solids}\ }\textbf {\bibinfo {volume} {57}},\ \bibinfo {pages} {355} (\bibinfo
  {year} {1983})}\BibitemShut {NoStop}%
\bibitem [{\citenamefont {Newman}\ \emph {et~al.}(2001)\citenamefont {Newman},
  \citenamefont {Strogatz},\ and\ \citenamefont {Watts}}]{newman2001random}%
  \BibitemOpen
  \bibfield  {author} {\bibinfo {author} {\bibfnamefont {M.~E.~J.}\
  \bibnamefont {Newman}}, \bibinfo {author} {\bibfnamefont {S.~H.}\
  \bibnamefont {Strogatz}},\ and\ \bibinfo {author} {\bibfnamefont {D.~J.}\
  \bibnamefont {Watts}},\ }\bibfield  {title} {\bibinfo {title} {Random graphs
  with arbitrary degree distributions and their applications},\ }\href
  {https://doi.org/10.1103/PhysRevE.64.026118} {\bibfield  {journal} {\bibinfo
  {journal} {Physical Review E}\ }\textbf {\bibinfo {volume} {64}},\ \bibinfo
  {pages} {026118} (\bibinfo {year} {2001})}\BibitemShut {NoStop}%
\bibitem [{\citenamefont {Molloy}\ and\ \citenamefont
  {Reed}(1995)}]{molloy1995}%
  \BibitemOpen
  \bibfield  {author} {\bibinfo {author} {\bibfnamefont {M.}~\bibnamefont
  {Molloy}}\ and\ \bibinfo {author} {\bibfnamefont {B.}~\bibnamefont {Reed}},\
  }\bibfield  {title} {\bibinfo {title} {A critical point for random graphs
  with a given degree sequence},\ }\href
  {https://doi.org/10.1002/rsa.3240060204} {\bibfield  {journal} {\bibinfo
  {journal} {Random Structures \& Algorithms}\ }\textbf {\bibinfo {volume}
  {6}},\ \bibinfo {pages} {161} (\bibinfo {year} {1995})}\BibitemShut {NoStop}%
\bibitem [{\citenamefont {Hiraoka}\ \emph {et~al.}(2016)\citenamefont
  {Hiraoka}, \citenamefont {Nakamura}, \citenamefont {Hirata}, \citenamefont
  {Escolar}, \citenamefont {Matsue},\ and\ \citenamefont
  {Nishiura}}]{hiraoka2016}%
  \BibitemOpen
  \bibfield  {author} {\bibinfo {author} {\bibfnamefont {Y.}~\bibnamefont
  {Hiraoka}}, \bibinfo {author} {\bibfnamefont {T.}~\bibnamefont {Nakamura}},
  \bibinfo {author} {\bibfnamefont {A.}~\bibnamefont {Hirata}}, \bibinfo
  {author} {\bibfnamefont {E.~G.}\ \bibnamefont {Escolar}}, \bibinfo {author}
  {\bibfnamefont {K.}~\bibnamefont {Matsue}},\ and\ \bibinfo {author}
  {\bibfnamefont {Y.}~\bibnamefont {Nishiura}},\ }\bibfield  {title} {\bibinfo
  {title} {Hierarchical structures of amorphous solids characterized by
  persistent homology},\ }\href {https://doi.org/10.1073/pnas.1520877113}
  {\bibfield  {journal} {\bibinfo  {journal} {Proceedings of the National
  Academy of Sciences}\ }\textbf {\bibinfo {volume} {113}},\ \bibinfo {pages}
  {7035} (\bibinfo {year} {2016})}\BibitemShut {NoStop}%
\bibitem [{\citenamefont {Jacobs}\ and\ \citenamefont
  {Thorpe}(1995)}]{jacobs1995}%
  \BibitemOpen
  \bibfield  {author} {\bibinfo {author} {\bibfnamefont {D.~J.}\ \bibnamefont
  {Jacobs}}\ and\ \bibinfo {author} {\bibfnamefont {M.~F.}\ \bibnamefont
  {Thorpe}},\ }\bibfield  {title} {\bibinfo {title} {Generic rigidity
  percolation: The pebble game},\ }\href
  {https://doi.org/10.1103/PhysRevLett.75.4051} {\bibfield  {journal} {\bibinfo
   {journal} {Physical Review Letters}\ }\textbf {\bibinfo {volume} {75}},\
  \bibinfo {pages} {4051} (\bibinfo {year} {1995})}\BibitemShut {NoStop}%
\bibitem [{\citenamefont {Jacobs}\ and\ \citenamefont
  {Thorpe}(1996)}]{jacobs1996}%
  \BibitemOpen
  \bibfield  {author} {\bibinfo {author} {\bibfnamefont {D.~J.}\ \bibnamefont
  {Jacobs}}\ and\ \bibinfo {author} {\bibfnamefont {M.~F.}\ \bibnamefont
  {Thorpe}},\ }\bibfield  {title} {\bibinfo {title} {Generic rigidity
  percolation in two dimensions},\ }\href
  {https://doi.org/10.1103/PhysRevE.53.3682} {\bibfield  {journal} {\bibinfo
  {journal} {Physical Review E}\ }\textbf {\bibinfo {volume} {53}},\ \bibinfo
  {pages} {3682} (\bibinfo {year} {1996})}\BibitemShut {NoStop}%
\bibitem [{\citenamefont {Laman}(1970)}]{laman1970}%
  \BibitemOpen
  \bibfield  {author} {\bibinfo {author} {\bibfnamefont {G.}~\bibnamefont
  {Laman}},\ }\bibfield  {title} {\bibinfo {title} {On graphs and rigidity of
  plane skeletal structures},\ }\href {https://doi.org/10.1007/BF01534980}
  {\bibfield  {journal} {\bibinfo  {journal} {Journal of Engineering
  Mathematics}\ }\textbf {\bibinfo {volume} {4}},\ \bibinfo {pages} {331}
  (\bibinfo {year} {1970})}\BibitemShut {NoStop}%
\bibitem [{\citenamefont {Chubynsky}\ and\ \citenamefont
  {Thorpe}(2007)}]{chubynsky2007}%
  \BibitemOpen
  \bibfield  {author} {\bibinfo {author} {\bibfnamefont {M.~V.}\ \bibnamefont
  {Chubynsky}}\ and\ \bibinfo {author} {\bibfnamefont {M.~F.}\ \bibnamefont
  {Thorpe}},\ }\bibfield  {title} {\bibinfo {title} {Algorithms for
  three-dimensional rigidity analysis and a first-order percolation
  transition},\ }\href {https://doi.org/10.1103/PhysRevE.76.041135} {\bibfield
  {journal} {\bibinfo  {journal} {Physical Review E}\ }\textbf {\bibinfo
  {volume} {76}},\ \bibinfo {pages} {041135} (\bibinfo {year}
  {2007})}\BibitemShut {NoStop}%
\bibitem [{\citenamefont {Neophytou}\ \emph {et~al.}(2024)\citenamefont
  {Neophytou}, \citenamefont {Starr}, \citenamefont {Chakrabarti},\ and\
  \citenamefont {Sciortino}}]{neophytou2024}%
  \BibitemOpen
  \bibfield  {author} {\bibinfo {author} {\bibfnamefont {A.}~\bibnamefont
  {Neophytou}}, \bibinfo {author} {\bibfnamefont {F.~W.}\ \bibnamefont
  {Starr}}, \bibinfo {author} {\bibfnamefont {D.}~\bibnamefont {Chakrabarti}},\
  and\ \bibinfo {author} {\bibfnamefont {F.}~\bibnamefont {Sciortino}},\
  }\bibfield  {title} {\bibinfo {title} {Hierarchy of topological transitions
  in a network liquid},\ }\href {https://doi.org/10.1073/pnas.2406890121}
  {\bibfield  {journal} {\bibinfo  {journal} {Proceedings of the National
  Academy of Sciences}\ }\textbf {\bibinfo {volume} {121}},\ \bibinfo {pages}
  {e2406890121} (\bibinfo {year} {2024})}\BibitemShut {NoStop}%
\end{thebibliography}
\end{document}